# Thermodynamic Explanation for the Cosmic Ubiquity of Organic Pigments

Karo Michaelian[a] and Aleksandar Simeonov[b]
a) Instituto de Física, UNAM, Circuito Interior de la Investigación Científica, Cuidad Universitaria, México D.F., Mexico, C.P. 04510.
b) Independent researcher, Bigla str. 7, Skopje, The former Yugoslav Republic of Macedonia.

**Abstract.** There is increasingly more evidence being accumulated for the occurrence of large amounts of organic material in the cosmos, particularly in the form of aromatic compounds. These molecules can be found on the surface of Earth and Mars, in the atmospheres of the larger planets and on many of their satellites, on asteroids, comets, meteorites, the atmospheres of red giant stars, interstellar nebulae, and in the spiral arms of galaxies. Many of these environments are expected to be of low temperature and pressure, implying that the Gibb's free energy for the formation of these complex molecules should be positive and large, suggesting that their existence could only be attributed to non-equilibrium thermodynamic processes. In this article we first review the evidence for the abundance of these molecules in the cosmos and then describe how the ubiquity can be explained from within the framework of non-equilibrium thermodynamics on the basis of the catalytic properties of these pigment molecules in dissipating photons of the ultraviolet and visible emission spectra of neighboring stars, leading to greater local entropy production. A relation between the maximum wavelength of absorption of these organic pigments and the corresponding stellar photon environment, provides a guide to determining which aromatic compounds are most probable in a given stellar neighborhood, a postulate that can be verified on Earth. It is suggested that at least some of the baryonic dark matter may be associated with these molecules which emit in the extreme infrared with many, but weak, emission lines, thus so far escaping detection. This thermodynamic explanation for the ubiquity of these organic molecules also has relevance to the possibility of life, both as we know it, and as we may not know it, throughout the universe.

# 1. Introduction

In 1937, the detection of the methylidyne (CH) radical in early type stars (O, B and A stars) (Dunham, 1937; Swings and Rosenfeld, 1937) ushered in an era of astrochemical discoveries. It was simultaneously the first molecule and the first organic compound to be ever discovered in extrasolar space.

An upsurge of molecular discoveries followed during the sixties and seventies, with the discovery of ammonia (Cheung et al., 1968), water (Cheung et al., 1969), formaldehyde (Snyder et al., 1969), methanol (Ball et al., 1970), carbon monoxide (Wilson et al., 1970). Of great importance to the question of the origin of life was the detection of N-containing organics like hydrogen cyanide (Snyder and Buhl, 1971), methylamine (Kaifu et al., 1974), and cyanamide (Turner et al., 1975), since it had been shown that HCN in the presence of water and a free energy source is sufficient to produce the DNA purines and pyrimidines (Oró, 1960; Oró and Kimball, 1961; Sanchez et al., 1967; Ferris et al., 1968). It therefore came as no surprise when the nucleobases were discovered in carbonaceous meteorites, along with some unusual and terrestrially rare analogs (purine, 2,6-diaminopurine, and 6,8-diaminopurine), indicative of their extraterrestrial origin (Stocks and Schwartz, 1981; Callahan et al., 2011). Another important discovery was that of interstellar acetylene (Lacy et al., 1989) a precursor molecule to benzene and hence other aromatic and polyaromatic organic molecules (Cernicharo et al., 2001).

As of 2016, approximately 200 gas-phase molecular species, most of them organic, have been recorded as common constituents of the interstellar medium and circumstellar shells (Kwok, 2016; McGuire et al., 2016).

Pivotal to the search and the accumulation of data has been the development of millimeter-wave and infrared spectroscopy which, through the detection of the stretching and bending vibrational modes of covalent bonds, as well as rotational transitions, has confirmed the existence of a vast array of these molecular types in space (Hearnshaw, 2014).

Heavier molecules, like organic polymers, have long since been speculated to exist in interstellar space. The relative abundance of the element carbon (forth most abundant element in the universe after H, He and O – Clayton, 2003), its unique covalent bonding flexibility and the abovementioned pioneering discoveries of low molecular weight organics, which can act as chemical precursors to other molecules, have spurred the search for larger, more complex organics in space.

In some unconventional and truly startling comparisons, Fred Hoyle and Chandra Wickramasinghe, the founders of the theory of panspermia, showed how the extinction spectra over the ultraviolet and visible of interstellar nebulae can be fitted well by the absorption spectra of freeze dried algae and bacteria (Hoyle and Wickramasinghe, 1979). More recent higher resolution data refute the quality of the fit (Pendleton and Allamandola, 2002), but there can be little doubt that the complexity of the absorbing material in the interstellar clouds must be somehow analogous to that of the biological macromolecules. The Extended Red Emission (ERE) feature, for example, which is a photoluminescence process powered by UV photons with a broad peak emission wavelength around 650-800 nm (Witt and Boroson, 1990), most closely parallels the optical behavior of chlorophyll and other porphyrinic bio-pigments, that have absorbance in the near-UV/blue and fluorescence in the red (Hoyle and Wickramasinghe, 1999).

Moreover, the ERE is only one of a family of universal spectroscopic features, which, although not yet fully explained, are typically regarded as spectral "fingerprints" of complex organic compounds throughout the cosmos. These include: the Unidentified Infrared Emission (UIE) bands, the Diffuse Interstellar absorption Bands (DIBs), the 217 nm feature, and the 21 and 30 micron emission features (Kwok, 2012 and references therein).

In particular, the UIE bands, which are emission bands originating from numerous galactic and extragalactic astronomical environments, have traditionally been considered as the spectral signatures of polycyclic aromatic hydrocarbons (Allamandola et al., 1985, 1989), although newer spectroscopic analyses (Kwok and Zhang, 2011, 2013) revive earlier proposals (Papoular et al., 1989) for even more complex chemical structure of the carriers of these bands; very reminiscent of the complex macromolecular structure of kerogen and coal on Earth, and the Insoluble Organic Matter (IOM) of carbonaceous chondrites (Kerridge, 1999).

There is thus unequivocal evidence for the formation of large and complex organic molecules throughout the universe. However, these complex molecules are prone to photo-ionization and destruction by charged particles, x-rays, and the same high-frequency UV photons that probably formed them. Therefore, their ubiquity, and thus the probable existence of life as we know it in the cosmos, requires an explanation that goes beyond near-equilibrium reaction pathways of formation. That these molecules are the result of a nonequilibrium thermodynamic processes can be seen from the fact that their occurrence is dependent upon a continuous flow of UV photons from the neighboring star. However, this paper goes further to show that the ubiquity of organic pigments throughout the cosmos, wherever there exists UV light interacting with material, is a result of the non-linear, non-equilibrium nature of the photochemical process leading to the organic molecules. Specifically, it is shown that these molecules can be considered as microscopic self-organized dissipative structures which form and proliferate to dissipate the generalized thermodynamic force; the stellar photon potential formed between the high energy photon outflow of a neighboring star and the low energy (microwave) photon cosmic background of space. These pigment molecules owe their existence and proliferation to the entropy production that they perform by dissipating high energy photons into heat.

## 2. Evidence for the cosmic ubiquity of organic pigments

Infrared observations by Gillett et al. (1973) and Russell et al. (1977, 1978) revealed that the IR spectra of bright astronomical objects with associated gas and dust particles, such as H$_{II}$ regions and planetary nebulae, are dominated by strong, broad, and resolved emission bands at 3.3, 6.2, 7.7, 8.6, and 11.3 μm. Since the carrier of the bands remained unidentified for almost a decade, they were dubbed the Unidentified Infrared Emission (UIE or UIR) bands. The fact that their intensity showed a direct correlation with carbon abundance within the source, naturally implied a carbon-based carrier.

Knacke (1977) and Duley and Williams (1981) first suggested aromatic carriers, since the bands peak at wavelengths corresponding to the stretching and bending modes of various CH and CC bonds in aromatic hydrocarbons, so they also became known as the Aromatic Infrared Bands (AIB or AIR).

With the launch of the *Infrared Space Observatory (ISO)* mission in 1995 the UIE features were observed with better spectral resolution, covering a broader spectral range, and in a larger sample of sources. They were found to dominate the mid-IR emission of almost all astronomical objects except for asymptotic giant branch (AGB) stars and deeply embedded young stellar objects (Tielens, 2005). It was estimated that around 20-30% of the infrared radiation of our galaxy is being emitted in these infrared emission bands (Mattila et al., 1999), and up to 20% of the total radiation output of distant starburst galaxies (Smith et al., 2007; Groves et al., 2008), indicating that their carrier must be an extremely common and abundant cosmic material.

Solar System objects, such as interplanetary dust particles (IDPs), carbonaceous meteorites, micrometeorites, Martian rocks (Kwok, 2009), comets (Keller et al., 2006), Titan's upper atmosphere (López-Puertas et al., 2013) are also known to display these infrared features.

A wide variety of hydrocarbon and carbonaceous materials comprising aromatic units have been proposed as possible carriers of the AIB bands, including; hydrogenated amorphous carbon (HAC), quenched carbonaceous composites (QCC), polycyclic aromatic hydrocarbons (PAHs), soot and carbon nanoparticles, fullerenes and fullerene-like particles, coal and kerogen, petroleum fractions etc (Henning and Salama, 1998). Among all these models, the PAH hypothesis developed by Leger and Puget (1984) and Allamandola et al. (1985, 1989) has gained most attention, therefore the UIE bands are usually referred to in the literature simply as the PAH bands.

Polycyclic aromatic hydrocarbons (PAHs) are benzene rings of *sp²*-hybridized C-atoms linked to each other in a plane, with H atoms or other functional groups saturating the outer bonds of peripheral C-atoms. Electron delocalization over their entire carbon skeleton makes them exceptionally stable because photonic excitation energy quickly becomes dispersed over the entire molecule before it can break a single bond.

According to the most accepted hypothesis, the infrared bands are attributed to the vibrational relaxation of PAH molecules containing around 50 C-atoms on average, which are stochastically heated to high temperatures (~1000 K) by the absorption of individual high-frequency UV photons (Puget and Leger, 1989; Tielens, 2008). In the low density environment of the interstellar medium, where collisional de-excitation is not possible, these UV-pumped, gas-phase PAH molecules undergo spontaneous de-excitation via infrared fluorescence which gives the AIB features.

While the hypothesis requires strong UV radiation field to excite the PAH molecules (PAHs absorb strongly only in the UV), Uchida et al. (1998, 2000) observed UIE features in the colder radiation environments of reflection nebulae, implying that other types of molecules may be responsible at least for some part of the emissions.

In support of this supposition is the fact that the PAH model is unable to explain the appearance of some aliphatic emission features at 3.4, 6.9 and 7.2 μm and the superimposition of the AIBs upon broad and strong emission plateau features at 6–9, 10–15, and 15-20 μm, which were shown to be of aliphatic nature (Ehrenfreund et al., 1991; Kwok et al., 2001).

Kwok and Zhang (2011, 2013) made spectral decomposition analyses of archival spectroscopic observations of the UIE bands and showed that the data are most consistent not with pure, free-flying PAH molecules, but with solid organic nanoparticles that have a mixed aromatic–aliphatic ($sp^2$-$sp^3$) structure. They called this new model MAON (Mixed Aromatic/Aliphatic Organic Nanoparticle). The proposed MAON emission carriers are quite different from the pure hydrocarbon, planar, gas-phase, relatively small molecular (~50 C-atoms) PAHs, where the structure, no matter how large, is regular with repeatable patterns. MAONs are amorphous solids of hundreds or thousands of C-atoms and impurity elements such as O, N, and S, with a disorganized, three-dimensional macromolecular structure comprised of aromatic and aliphatic units, each with variable sizes and random orientations.

Since the late 1980's Renaud Papoular et al. have argued that the observed spectral properties of the UIE bands most closely resemble those of coal and kerogen (Papoular et al., 1989; Guillois et al. 1996; Papoular, 2001), which are actually organic, amorphous, macromolecular materials with randomly oriented clusters of aromatic rings linked by long aliphatic chains, with O, N, S functional groups and heterocycles in their structure. This complex macromolecular structure is also very similar to the structure of the Insoluble Organic Matter (IOM) found in the Murchison meteorite and identified in laboratory analyses by Derenne and Robert (2010). More than 70% of the organic matter in carbonaceous chondrites, such as the Murchison, Murray, Tagish Lake, and Orgueil meteorites is in this form of insoluble amorphous solids (Cronin et al., 1987, Sephton, 2002). The structure of the IOM in all these meteorites shows remarkable resemblance (Cody and Alexander, 2005), while isotopic ratios point to their interstellar origin (Nakamura-Messenger et al., 2006; Martins et al., 2008).

It is interesting that in the UV-poor environment of protoplanetary nebulae (PPN), the aliphatic chains of these UIE emitters constitute a significantly larger percentage of the entire macromolecular structure when compared to the UV-intense regions of planetary nebulae (PN) where the aromatic component is predominant (Kwok et al., 2001; Kwok, 2007). For example, in the PPN phase of stellar evolution, the 3.4 and 6.9 μm aliphatic emission features are as strong as the 3.3, 6.2, 7.7, and 11.3 μm aromatic emission features, but as the star evolves to the PN stage (a process of less than few thousand years), the aromatic features become considerably stronger (Kwok, 2007). This phenomenon has been explained as the result of photochemistry, where the onset of UV radiation modifies the aliphatic side groups through cyclizations and isomerizations, and transforms them into aromatic ring structures, making the macromolecule more aromatic (Joblin et al., 1996; Sloan et al., 2007).

In the following section, we will show how, under an imposed UV-C light flux from nearby stars, non-equilibrium thermodynamic principles based on the maximization of entropy production through UV-photon dissipation would indicate the formation, proliferation, and aromatization of these organic compounds.

Much of the diverse organic material found on the surface of Earth is in the form of pigments that absorb strongly across the UV and visible wavelength region and dissipate this energy rapidly into heat (vibrational energy) which is then further dispersed over the vibrational modes of the surrounding solvent water molecules (Michaelian, 2011, 2013; Michaelian and Simeonov, 2014, 2015). This is true of DNA and RNA which absorb and dissipate strongly from 220 to 280 nm, for the aromatic amino acids tyrosine, tryptophan, and phenylalanine which absorb strongly from 220 nm to 300 nm, for the mycosporine-like amino acids and scytonemin which absorb over 300 to 500 nm, and for the other common pigments of phototropic organisms; chlorophylls, phycobilins, carotenoids, anthocyanins, which absorb from the visible to the infrared.

It is an intriguing fact that most of the fundamental molecules of life (those found in all three domains of life) are also pigments which absorb in the UV-C and have chemical affinity to RNA and DNA. Through Förster-type energy transfer, as well as through the coupling of vibronic modes (Vekshin, 2005), these molecules can also use RNA and DNA as acceptor quencher molecules to dissipate extremely rapidly (through a conical intersection) the absorbed photon excitation energy into heat (Michaelian and Simeonov, 2014, 2015).

Planet Earth has for a long time been considered as the only body in the solar system that contains organic matter, usually attributed to the activities of life. The conventional picture for the chemical composition of other solar system bodies (planets, satellites, asteroids, comets, and dust particles) is that they are entirely made up of metals, minerals, ices and gases, with either no or only traces of organics. However, with the development of more sophisticated technologies like sending spacecrafts to solar system bodies for close spectroscopic observation and direct sample collection, organic substances are increasingly being recognized as a major component of planetary chemistry.

One of the first hints of the presence of organic matter on the surface of asteroids came from their deep red colors and low albedos (0.01-0.15). Such intense colors and low albedos are difficult to explain considering only inorganic minerals and ices but can easily be explained by organic pigments with a complex kerogen-like macromolecular composition (Gradie and Veverka, 1980). Roush and Cruikshank (2004) have noted that terrestrial dark pigmented organic solids like tar sands, asphaltite, pyrobitumen, and kerite, which all have complex aromatic/aliphatic composition, have very low albedos and dark reddish-brown colors, reminiscent of asteroids.

Saturn's largest moon Titan, the only know natural satellite to have a dense atmosphere, has attracted much attention since the Cassini–Huygens unmanned spacecraft, which arrived there in 2004, revealed that Titan's atmosphere contains haze-like solid particles that are the result of condensation of organics. The observations support the hypothesis that methane and nitrogen molecules excited by UV photons react to form polymeric hydrogenated carbon-nitride compounds, called tholins, that give the distinctive thick layer of orange-brown haze in Titan's lower stratosphere (Waite, 2007; Nguyen et al., 2007). Cassini RADAR observations found that these organic nanoparticles condense on surface sand grains that are blown into longitudinal dark-colored dunes by winds. Most interestingly, some of these organic nanoparticles end up dissolved in the numerous lakes and rivers of liquid methane and ethane in Titan's polar regions, which exhibit active liquid-gas phase cycling, similar to the water cycle on Earth, although at much lower temperature (Atreyaa et al., 2006).

The Cassini spacecraft has also found evidence of mixed aromatic/aliphatic hydrocarbons on other Saturnian satellites: Iapetus, Phoebe and Hyperion (Clark et al., 2005; Cruikshank et al., 2007, 2008). Iapetus is the third largest satellite of Saturn, locked in synchronous rotation about the planet, with the leading hemisphere and sides of dark brown color and low albedo (0.03–0.05), and most of the trailing hemisphere and poles of bright color and high albedo (0.5–0.6) (Denk et al., 2010). This difference in coloring between the two hemispheres is striking. Temperatures on the dark region's surface reach about 129 K at the equator, while the bright surfaces reach only about 113 K due to less sunlight absorption (Spencer and Denk, 2010). The low-albedo material exhibits mainly an aromatic hydrocarbon signature (Cruikshank et al., 2014).

Trans-Neptunian objects (TNO) are a group of minor bodies that orbit the Sun in the outer solar system beyond the planet Neptune. Photometric observations in the visible have found the colors of some of these objects to be intensely red (Jewitt and Luu, 2001), which can be an indication of the presence of organic material on their surface.

Exceptionally rich in organic content and organic diversity are comets (Mumma and Charnley, 2011). They represent agglomerates of rocky debris, frozen gases and ices, and are likely the most primitive bodies in the solar system (Whipple, 1978). Dust samples from the coma of comet Wild 2 were collected by the Stardust spacecraft and returned to Earth in 2006. The samples were analyzed with various techniques, giving evidence of rich organic content, but mostly in the form of aromatic-aliphatic macromolecules similar to the IOM of meteorites but with higher N and O content. Of particular significance is that the aromatic compounds are not pure hydrocarbons but have a very high N content, where N is incorporated predominantly in the form of aromatic nitriles (R–C≡N); a fact of great astrobiological implication as aromatic nitriles are precursory to many organic pigments and other fundamental biological molecules, and comets are likely to have contributed to Earth's prebiotic chemical inventory (Keller et al., 2006; Clemett et al., 2010).

## 3. A non-equilibrium thermodynamic explanation for organic pigment ubiquity

It is thus now abundantly clear that a large class of pigment molecules exist throughout the cosmos in the form of tholins, polyaromatic hydrocarbons (PAHs) and/or Mixed Aromatic/Aliphatic Organic Nanoparticles (MAONs) and that these systems are absorbing light in the UV and visible spectrums and dissipating it into infrared emission bands. The dissipative process in which a part of the incident solar spectrum is absorbed and dissipated into heat by organic pigment molecules at the Earth's surface is by far the most important entropy producing process occurring on Earth. Similarly, the interstellar reddening due to photon dissipation by the interstellar PAHs and MAONs is probably the most important entropy producing process occurring in the cosmos after nuclear fusion and the expansion of the universe.

It will now be shown that the formation of these pigment molecules can be viewed as microscopic self-organization of material (microscopic dissipative structuring) in response to the imposed UV photon potential from nearby stars and that pigment proliferation to well beyond expected equilibrium concentrations follows from the same non-linear, non-equilibrium thermodynamic principles. Our thermodynamic analysis will follow that of Prigogine (1967) for a set of purely autocatalytic chemical reactions, the difference being that instead of dissipating chemical potentials, our cosmic pigment system will be dissipating photochemical potentials related to the photon pressure of the relevant stellar spectra.

Advantages for microscopic dissipative structuring utilizing photochemical reactions over normal thermally induced chemical reactions occurring in the electronic ground state are various (König, 2015);

1. Reactions may occur that are very endothermic in the ground state since the absorbed photon donates its energy to the molecule. For example, the energy of a photon of 260 nm (UV-C) is 4.77 eV or 110 Kcal/mol.
2. In the electronic excited state antibonding orbitals are occupied and this may allow reactions which are not possible for electronic reasons in the ground state.
3. Photochemical reactions can involve singlet and triplet states. Thermal reactions usually only involve singlet states. Therefore, photochemical reaction intermediates may be formed which are not accessible under thermal conditions.

The coupled photochemical dissipative process involving UV-C light dissipation and cosmic pigment production, which drives pigment proliferation, is the transformation of a black-body spectrum at a high temperature, that of the surface of a nearby star (e.g. our sun at ~5760 K) into an intermediate black-body spectrum at a lower temperature (that of the vibrational spectrum of the low molecular weight precursor molecules of the pigments) after absorbing UV-C photons, and finally the dissipation to a black-body spectrum at a temperature of the infrared emission bands (~287 K). Besides this photon dissipation process, we also have a "parasitic" photochemical reaction producing pigments from the excited precursor molecules. These pigments then act as catalysts for the dissipation of the incident stellar spectrum into the infrared emission spectrum, thereby making the global dissipation process autocatalytic and this brings the non-linearity into the non-equilibrium thermodynamic process. It is this nonlinearity that leads to pigment proliferation.

We will show that in the thermodynamic stationary state, the concentration of the pigment molecules can grow to many orders of magnitude what its concentration would be expected if the system were closer to equilibrium or if the pigment were not an effective catalyst in the photon dissipation. This is the same reasoning which explains the concentration of a product catalyst in a normal chemical reaction growing to many orders its expected equilibrium concentration due to its involvement in the dissipation of the imposed chemical potential (Prigogine, 1967).

Photons carry momentum, and thus for a collection of photons in equilibrium (a black-body spectrum), a unique equilibrium pressure can be defined which is found to go as the fourth power of the temperature. It is assumed here that the rate of energy conversion between spectra will be linearly proportional to the difference of the photon pressures of the different spectra. This is similar to assuming that the rate of a chemical reaction is proportional to the difference in the concentrations of the reactant and product (with equilibrium and rate constants set equal to one), or that the flow of material is proportional to the gas pressure differences, or that the flow of electrical current is proportional to the difference in electrical potential. These latter two linear relationships are valid when the mean free paths (between scattering events) of the particles are small with respect to the size of the system, i.e. that the system attains local equilibrium, while for the former chemical reactions (or photon dissipation), it is only required that the reactants and products (or photons) retain a Maxwell-Boltzmann (or Boltzmann) distribution of the velocities (energies) of the particles (or photons) involved.

There are then four relevant photon pressures, 1) $P_S$ corresponding to the stellar photon spectrum arriving at the site of the precursor molecules, approximately black-body with a temperature equal to that of the surface of the local star, 2) $P_I$ corresponding to the vibrational, assumed black-body, temperature of the precursor molecules after absorbing the energy of a UV-C photon and then electronic de-excitation through a conical intersection into vibrational energy, and 3) $P_E$ the photon spectrum of the infrared emission bands, 4) $P_P$ corresponding to the spectrum emitted by the produced pigment which will be dependent on the concentration of the pigments and the energy diverted into forming their covalent bond structures.

Although using a black-body approximation is wrought with error since the incident radiation has particular directionality properties and only approximates a black-body spectrum, the purpose of the calculation here is to show only qualitatively how the proliferation of organic pigments in the cosmos can be explained from a non-linear irreversible thermodynamic analysis of the photon dissipation process. A more accurate derivation of this would employ Planck's formula for the entropy of an arbitrary beam of photons.

Employing the formalism of classical irreversible thermodynamics (Prigogine, 1967), the entropy production $P$ is a product of generalized forces, $X$, times generalized flows, $J$,

$$P = \frac{d_i S}{dt} = \sum_k J_k X_k \geq 0. \qquad (1)$$

As an example, for a chemical reaction, the generalized force $X$ is the affinity over the temperature $A/T$, where the affinity $A$ for the reaction is equal to the stoichiometric coefficients of the reactants multiplied by the chemical potentials minus the same of the products. The chemical potentials, in turn, depend on the concentrations of the reactants and products and the free energy differences of the reactant and product states. The generalized flow $J$ for the irreversible process of a chemical reaction is the rate of the reaction.

For our photon dissipation irreversible process to be considered here, the generalized forces are the photon spectrum affinities determined from the photon pressures, and the generalized flows are the flows of energy from the incoming short wavelength spectrum to the emitted long wavelength spectra.

The change of the entropy production $dP$ can be decomposed into two parts, one related to the change of forces and the other to the change of flows (Prigogine, 1967),

$$dP = d_X P + d_J P = \sum_k J_k dX_k + \sum_k X_k dJ_k \tag{2}$$

where the sums are over all $k$ irreversible processes occurring within the system.

In the whole domain of the validity of thermodynamics of irreversible processes, and under constant external constraints, the contribution of the time change of the forces to the entropy production is negative or zero,

$$d_X P \leq 0. \tag{3}$$

This is known as the general (or universal) evolutionary criterion, or the Glansdorff-Prigogine criterion, since it was first established by Glansdorff and Prigogine (1964) (see also Prigogine (1967)). The general evolutionary criterion is the most general result that has thus far been obtained from Classical Irreversible Thermodynamics. For systems with constant external constraints (for example, a constant flux of UV photons), the system will eventually come to a thermodynamic stationary state in which case (Prigogine, 1967),

$$d_X P = 0. \tag{4}$$

With this background in classical irreversible thermodynamics, we can now analyze our particular irreversible processes consisting of the conversion of energy through different photon spectra. First, the incident flow of energy of the photon spectrum coming from the surface of the star $I_S(\lambda)$, which can be approximated by a black-body spectrum for convenience, $I_S(T_S)$, where $T_S$ is the temperature of the surface of the star (say 5760 K, for a G-type star like our sun), is converted, by absorption on precursor molecules followed by either a photochemical reaction leading to a pigment molecule, or by the dissipation directly into heat giving an intermediate spectrum of intermediate temperature (also assumed to be black-body), $I_I(T_I)$, with say $T_I = 1000$ K (corresponding roughly to the vibrational temperature of a molecule of 50 atoms or so after absorbing a single UV-C photon). This spectrum is then converted through interaction of the molecule with its gas or solid MAON surroundings into the red-shifted photon spectrum of the infrared emission bands $I_E(T_E)$, with say $T_E = 287$ K. Note that in the very low density environment of space, aggregation of individual molecules into large structures like MAONs would be thermodynamically selected since this would allow distribution of the incident photon energy over more microscopic degrees of freedom, or, in other words, emission at lower frequencies, giving greater entropy production. Isolated molecules in a vacuum would have to fluoresce at high frequencies in order to dissipate the excitation energy of a single UV-C photon, leading to less entropy production.

This process can be equivalently looked at as one in which, at the intermediate temperatures of the vibrationally excited precursor molecules, a photochemical reaction may take place in which some of the free energy of the vibrationally excited molecules is diverted into the production of organic pigments leading to a emitted photon spectra of the newly formed pigments which is dependent on the concentration C of the pigments and how much free energy went into forming the relevant covalent bonding and is also assumed to be characterized by a black-body spectrum of temperature $T_P$.

These newly formed pigments themselves act as catalysts for the overall dissipation process of $I_S(T_S) \to I_E(T_E)$, and can be even more catalytic by, for example, providing local heat of dissipation to increase the probability of the photochemical reaction leading to a pigment $I_I(T_I) \to I_P(T_P)$. These two effects make the dissipation process auto-catalytic. A schematic diagram for this auto-catalytic photochemical process can be given as follows,

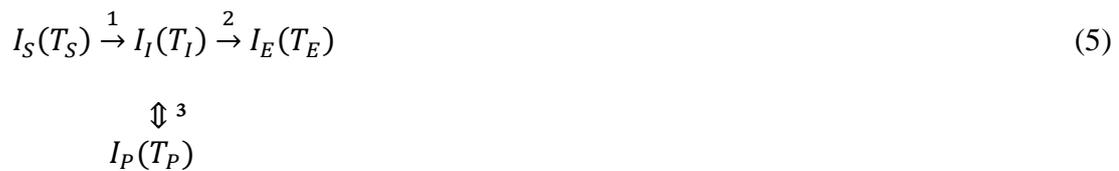

$$I_S(T_S) \xrightarrow{1} I_I(T_I) \xrightarrow{2} I_E(T_E) \quad\quad\quad (5)$$

$$\Updownarrow 3$$
$$I_P(T_P)$$

Since the system is far from equilibrium, the backward rate constants for the spectra conversions can be considered as being essentially zero, which is also the case for the photochemical reactions leading to the pigments (reaction 3). Note, however, that if a high energy photon can produce a pigment, a similar photon could also destroy it. There is escape from this predicament if the pigment is fortuitously produced with a conical intersection allowing rapid non-radiative dissipation, thereby neutralizing the destructive potential of the UV-C photons. Any pigments being produced without a conical intersection, but having chemical affinity to a molecule that has one, so that the pair could operate in the donor-quencher mode, would also be spared from the destructive potential of the UV-C photons. Such a selection process would tend to build up pigments with conical intersections or pigments with chemical affinity to those that have one, perhaps forming complexes like the MAON structures.

Since the photon spectra are all considered to be black-body, they are completely characterized by one variable which is the black-body temperature $T$, or equivalently by the photon pressure $P$ which goes as $T^4$. Thus, the conversion of the energy through the different spectra and the energy flow into the formation of the pigments can be characterized, in a first approximation, by the pressures corresponding to the assumed black-body spectra. We can therefore write equation (5) alternatively in terms of pressures,

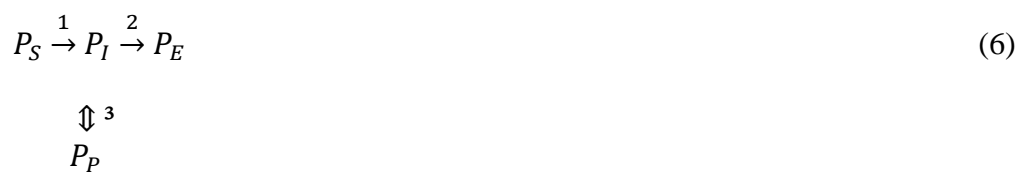

$$P_S \xrightarrow{1} P_I \xrightarrow{2} P_E \quad\quad\quad (6)$$

$$\Updownarrow 3$$
$$P_P$$

where, as mentioned above, $P_P$ is the photon "pressure" related to the black-body temperature of the pigments after production which is related to their concentration. It is also assumed that the rate of production of the organic pigments (reaction 3) is linearly related to the photon pressure difference $P_I - P_P$. This pressure difference is essentially related to the free energy available for the formation of the pigment. If the photon pressure $P_P$ associated with the newly formed pigments is low, then more photon energy will flow into the production process, pigments will be produced at a greater rate.

In terms of the affinities and flows of the different processes, we can write Prigogine's general evolutionary criterion, Eq. (4) once arriving at the stationary state, in the following form,

$$d_X P = dP - d_J P = d\left(\sum_k \frac{A_k}{T} v_k\right) - \sum_k \frac{A_k}{T} dv = 0, \tag{7}$$

where $A_1$ is the affinity for the conversion of energy of the solar photon spectrum into energy of the vibrationally excited intermediate precursor molecule photon spectrum, $A_2$ is the affinity for the conversion of energy of the excited intermediate precursor molecule photon spectrum into energy of the emitted photon spectrum in the infrared, and $A_3$ is the affinity for the photochemical reaction producing the organic pigment.

For the case of equilibrium photon distributions (black-body spectra), the affinities will go as the logarithm of the ratio of the photon pressures (Herrmann and Würfel, 2005),

$$A_1 = kT_I \log\left(\frac{P_S}{P_I}\right), \quad A_2 = kT_E \log\left(\frac{P_I}{P_E}\right), \quad A_3 = kT_P \log\left(\frac{P_I}{P_P}\right) \tag{8}$$

The $v_k$ in Eq. (7) are the rates of the corresponding energy conversion (dissipation) processes which we assume are related to the differences of the photon pressures (see discussion at the beginning of this section) attributed to the different black-body spectra. Since the organic pigments are assumed to act as catalysts for the conversion of energy from the stellar spectrum to the emitted spectrum (i.e. the local heat of photon dissipation by the pigment can be dissipated into the environment (reaction 1+ 2) or can act as a catalyst for the photochemical production of a new pigment (reaction 3)), and since $P_P$ is related to the concentration of the pigments, then we could model the autocatalytic process by multiplying the first dissipation process, $P_S \to P_I$, by a factor $(1+\alpha P_P)$ where α represents the effectiveness of the organic pigment as a catalyst for photon energy conversion to heat (i.e. α→∞ for an excellent catalyst, and α→0 for a completely ineffective catalyst). Therefore, the rates of conversion, assuming all constants of proportionality equal to one for convenience (again, we are only interested in showing qualitatively the non-equilibrium dynamics of pigment proliferation), are given by

$$v_1 = (1 + \alpha P_P)(P_S - P_I) \quad v_2 = (P_I - P_E) \quad v_3 = (P_I - P_P) \tag{9}$$

Note the non-linear relation between the forces, Eq. (9), and flows, Eq. (8).

Using Eq. (7) for the steady state together with Eqs. (8) and (9), taking the Boltzmann constant $k = 1$ for convenience, and observing that the free forces can be characterized in terms of the two free pressures, $P_I$ and $P_P$ (since $P_S$ and $P_E$ are fixed and given by the local star surface temperature to the fourth power and the infrared emission temperature to the fourth power respectively) gives

$$\frac{\partial}{\partial P_I}\left[(1+\alpha P_P)(P_S - P_I)\log\frac{P_S}{P_I} + (P_I - P_E)\log\frac{P_I}{P_E} + (P_I - P_P)\log\frac{P_I}{P_P}\right]$$
$$+ (1+\alpha P_P)\log\frac{P_S}{P_I} - \log\frac{P_I}{P_E} - \log\frac{P_I}{P_P} = 0 \qquad (10)$$

$$\frac{\partial}{\partial P_P}\left[(1+\alpha P_P)(P_S - P_I)\log\frac{P_S}{P_I} + (P_I - P_E)\log\frac{P_I}{P_E} + (P_I - P_P)\log\frac{P_I}{P_P}\right]$$
$$- \alpha(P_S - P_I)\log\frac{P_S}{P_I} + \log\frac{P_I}{P_P} = 0 \qquad (11)$$

At the steady state, we have (see Eq. (4)),

$$d_X P = \sum_k v_k dA_k = 0 \qquad (12)$$

which, after algebra, gives

$$v_1 = v_2, \quad v_3 = 0 \qquad (13)$$

which implies that at the steady sate no free energy is being diverted into further pigment production $v_3 = 0$.

Also, at the steady state, solving Eqs. (10) and (11) gives,

$$P_P = P_I = \frac{1}{2\alpha}\left[\alpha P_S - 2 + \left[4 + 4\alpha P_S(1-\gamma) + \alpha^2 P_S^2\right]^{1/2}\right]$$
$$\to (1/2)(P_S + P_E) \quad \text{for } \alpha \to 0$$
$$\to P_S \quad \text{for } \alpha \to \infty \qquad (14)$$

where we define $(1-\gamma) = P_S / P_E$ ($\gamma$ is, therefore, a measure of the "distance" from equilibrium of the system). Therefore, since $P_S$ is much greater than $P_E$ (the pressures go as the temperature to the fourth power for black-body spectra) equation (14) indicates the photon pressure related to the organic pigments $P_P$, or in other words the pigment concentration, or the free energy which has gone into pigment production, has increased due to its catalytic activity in forming new pigments and dissipating the stellar photon spectrum into the emitted spectrum.

The entropy production of the energy conversion processes, including catalytic activity of the organic pigment, is given by flows times forces,

$$\frac{d_i S}{dt} = \sum_k v_k \frac{A_k}{T_k} = (P_S - P_I)(1+\alpha P_P)\log\frac{P_S}{P_I} + (P_I - P_E)\log\frac{P_I}{P_E} + (P_I - P_P)\log\frac{P_I}{P_P} \qquad (15)$$

It can also be shown that the entropy production at the stationary state shifts to larger values as a result of the catalytic activity (see Prigogine, 1967, for the corresponding case of chemical reactions).

These results give the non-linear irreversible thermodynamic explanation for the proliferation of organic pigments in the cosmos. Their concentration can become much greater than that expected under near equilibrium conditions due to their catalytic nature in dissipating the stellar spectrum, depending on the ratio of $P_S / P_E = \left(\dfrac{T_S}{T_E}\right)^4$ which is a measure of the distance of the system from equilibrium.

In referring to purely chemical reactions, Prigogine (1967), in fact, noticed that such a result sheds light on the problem of the occurrence of complicated biological molecules in steady state concentrations which are of orders of magnitude larger than the equilibrium concentrations. In his 1967 book "Thermodynamics of Irreversible Processes" (Prigogine, 1967) he states, "Thus, for systems sufficiently far from equilibrium, kinetic factors (like catalytic activity) may compensate for thermodynamic improbability and thus lead to an enormous amplification of the steady state concentrations of the catalyst. Note that this is a strictly non-equilibrium effect. Near equilibrium, catalytic action would not be able to shift in an appreciable way the position of the steady state."

## 4. Summary and Discussion

The cosmos is replete with organic pigments which absorb at the wavelengths of prominence of the local star and dissipate these high energy photons into the infrared. The low probability of formation of these pigments under the low temperature and high vacuum conditions of space suggest that they are the result of non-equilibrium thermodynamic processes. In this article, we have suggested that these pigments should be considered as microscopic dissipative structures formed as a non-equilibrium thermodynamic response to dissipate the impressed stellar photon potential.

The proliferation of the pigments to concentrations many orders of magnitude greater than what could be expected under near equilibrium conditions can be explained within this non-equilibrium thermodynamic framework by invoking the autocatalytic nature of the pigments in dissipating the same photon potential that produced them.

Evidence in favor of our proposition is many fold;
1. Organic pigment ubiquity would not be expected under near equilibrium conditions.
2. Aromatic pigments are found to be more prevalent under environmental light conditions of higher energy UV photons, whereas aliphatic chain pigments are found under conditions of lower energy UV photons. That is, there is a direct relation between the UV flux from a local star and the maximum energy of absorption of the produced pigments (aromatic pigments absorb higher energy photons than aliphatic pigments).
3. The aggregation of the organic pigment molecules into MAON nano-sized structures can thermodynamically be explained by the greater dissipation provided by these structures with many more microscopic degrees of freedom for distributing the vibrational energy as compared to the isolated PAH's.
4. Proliferation of pigments by the same mechanism on Earth has led to the proliferation of the fundamental molecules of life, almost all of which dissipate in the UV-C range, exactly over a UV-C window which existed in Earth's atmosphere during the Archean (Michaelian 2013, Michaelian and Simeonov, 2015).

It may be that in regions of the cosmos, dissipative self-organization of material under stellar photon fluxes has evolved to the point where the emitted spectrum is so far towards the infrared such that the emission lines have become many and too weak to be detectable. It may thus be possible that some of this dissipative structured material is contributing to the baryonic dark matter. Dark matter has been determined to make up some 27% of the mass-energy density of the universe and baryonic dark matter makes up approximately 10% of this dark matter density if the universe is at the critical density for closure, and if the standard model for big bang nucleosynthesis is correct. Since the known baryonic matter makes up less than 5% of the mass-energy density of the universe (NASA, 2016), the amount of material structuring dissipatively into this type of baryonic dark matter could exceed 50% of all observable matter in the universe. Other explanations for the invisible baryonic dark matter such as supermassive black holes or MACHOS have difficulty in accounting for all of the missing mass.

If such self-organization into microscopic photon dissipative structures is indeed occurring in the cosmos, then it could be expected that the fundamental molecules of life (those found in all three domains of life), which appear to have been self-organized dissipative structures in the UV-C (Michaelian and Simeonov, 2015), should be ubiquitous throughout the cosmos. This would imply that the origin of life, as we know it (Michaelian, 2009; 2011) based on these molecules, should also be very common throughout the universe. Other forms of biotic-abiotic biospheres, such as the organic molecules floating in the clouds of Jupiter, absorbing sunlight and channeling the heat of dissipation into the multiple vortices and storms of the Jovian atmosphere, or the organic molecules in the methane lakes of Titan fomenting the methane rain cycle, may be equally, or even more, common.

**Acknowledgements:**
K.M. would like to thank DGAPA-UNAM project number IN-102316 for financial support.

**References**:

Allamandola, L.J., Tielens, A.G.G.M., and Barker, J.R., 1985. Polycyclic aromatic hydrocarbons and the unidentified infrared emission bands: auto exhaust along the Milky Way!. The Astrophysical Journal. 290, L25-L28.

Allamandola, L.J., Tielens, A.G.G.M., and Barker, J.R., 1989. Interstellar polycyclic aromatic hydrocarbons - The infrared emission bands, the excitation/emission mechanism, and the astrophysical implications. The Astrophysical Journal. 71, 733-775.

Atreyaa, S.K., Adamsa, E.Y., Niemann, H.B., Demick-Montelara, J.E., Owen, T.C., Fulchignoni, M., Ferri, F., Wilson, E.H., 2006. Titan's methane cycle. Planetary and Space Science. 54 (12), 1177.

Ball, J.A., Gottlieb, C.A., Lilley, A.E., and Radford, H.E., 1970. Detection of methyl alcohol in Sagittarius. The Astrophysical Journal. 162, L203-L210.

Callahan, M.P., Smith, K.E., Cleaves, H.J., Ruzicka, J., Stern, J.C., Glavin, D.P., House, C.H., Dworkin, J.P., 2011. Carbonaceous meteorites contain a wide range of extraterrestrial nucleobases. Proceedings of the National Academy of Sciences. 108 (34), 13995-13998.

Cernicharo, J., Heras, A.M., Tielens, A.G.G.M., Pardo, J.R., Herpin, F., Guelin, M., and Waters, L.B.F.M., 2001. *Infrared Space Observatory's* Discovery of $C_4H_2$, $C_6H_2$, and Benzene in CRL 618. The Astrophysical Journal. 546, L123-L126.

Cheung, A.C., Rank, D.M., Townes, C.H., Thornton, D.D., and Welch, W.J., 1968. Detection of $NH_3$ molecules in the interstellar medium by their microwave emission. Physical Review Letters. 21, 1701-1705.

Cheung, A.C., Rank, D.M., Thornton, D.D., and Welch, W.J., 1969. Detection of water in interstellar regions by its microwave radiation. Nature. 221, 626-628.

Clark, R.N., Brown, R.H., Jaumann, R., Cruikshank, D.P., Nelson, R.M., Buratti, B.J., McCord, T.B., Lunine, J., Baines, K.H., Bellucci, G., Bibring, J.P., Capaccioni, F., Cerroni, P., Coradini, A., Formisano, V., Langevin, Y., Matson, D.L., Mennella, V., Nicholson, P.D., Sicardy, B., Sotin, C., Hoefen, T.M., Curchin, J.M., Hansen, G., Hibbits, K., Matz, K.D., 2005. Compositional maps of Saturn's moon Phoebe from imaging spectroscopy. Nature. 435 (7038), 66-69.

Clayton, D., 2003. Handbook of Isotopes in the Cosmos: Hydrogen to Gallium, Cambridge Planetary Science, Cambridge University Press, Cambridge United Kingdom.

Clemett, S.J., Sanford, S.A., Nakamura-Messenger, K., Hörz, F., and McKay, D.S., 2010. Complex aromatic hydrocarbons in Stardust samples collected from comet 81P/Wild 2. Meteoritics and Planetary Science. 45, 701-722.

Cody, G.D. and Alexander, C.M.O'D., 2005. NMR studies of chemical structural variation of insoluble organic matter from different carbonaceous chondrite groups. Geochimica et Cosmochimica Acta. 69, 1085-1097.

Cronin, J.R., Pizzarello, S., and Frye, J.S., 1987. $^{13}$C NMR spectroscopy of the insoluble carbon of carbonaceous chondrites. Geochimica et Cosmochimica Acta. 51, 299-303.

Cruikshank, D.P., Dalton, J.B., Dalle Ore, C.M., Bauer, J., Stephan, K., Filacchione, G., Hendrix, A.R., Hansen, C.J., Coradini, A., Cerroni, P., Tosi, F., Capaccioni, F., Jaumann, R., Buratti, B.J., Clark, R.N., Brown, R.H., Nelson, R.M., McCord, T.B., Baines, K.H., Nicholson, P.D., Sotin, C., Meyer, A.W., Bellucci, G., Combes, M., Bibring, J.P., Langevin, Y., Sicardy, B., Matson, D.L., Formisano, V., Drossart, P., Mennella, V., 2007. Surface composition of Hyperion. Nature. 448 (7149), 54-56.

Cruikshank, D.P., Wegryn, E., Dalle Ore, C.M., Brown, R.H., Bibring, J. –P., Buratti, B.J., Clark, R.N., McCord, T.B., Nicholson, P.D., Pendleton, Y.J., Owen, T.C., Filacchione, G., Coradini, A., Cerroni, P., Capaccioni, F., Jaumann, R., Nelson, R.M., Baines, K.H., Sotin, C., Bellucci, G., Combes, M., Langevin, Y., Sicardy, B., Matson, D.L., Formisano, V., Drossart, P., and Mennella, V., 2008. Hydrocarbons on Saturn's satellites Iapetus and Phoebe. Icarus. 193, 334-343.

Cruikshank, D.P., Dalle Ore, C.M., Clark, R.N. and Pendleton, Y.J., 2014. Aromatic and aliphatic organic materials on Iapetus: Analysis of Cassini VIMS data. Icarus. 233, 306-315.

Denk, T., Neukum, G., Roatsch, T., Porco, C.C., Burns, J.A., Galuba, G.G., Schmedemann, N., Helfenstein, P., Thomas, P.C., Wagner, R.J., West, R.A., 2010. Iapetus: unique surface properties and a global color dichotomy from Cassini imaging. Science. 327 (5964), 435-439. doi: 10.1126/science.1177088. Epub 2009 Dec 10.

Derenne, S. and Robert, F., 2010. Model of molecular structure of the insoluble organic matter isolated from Murchison meteorite. Meteoritics and Planetary Science. 45, 1461–1475.

Duley, W.W. and Williams, D.A., 1981. The infrared spectrum of interstellar dust: surface functional groups on carbon. Monthly Notices of the Royal Astronomical Society. 196, 269-274.

Dunham, T.J., Jr., 1937. Interstellar neutral potassium and neutral calcium. Publications of The Astronomical Society of the Pacific. 49, 26-28.

Ehrenfreund, P., Robert, F., d'Hendencourt, L., and Behar, F., 1991. Comparison of interstellar and meteoric organic matter at 3.4 μm. Astronomy and Astrophysics. 252, 712-717.

Ferris, J.P., Sanchez, R.A., and Orgel, L.E., 1968. Studies in prebiotic synthesis. III. synthesis of pyrimidines from cyanoacetylene and cyanate. Journal of Molecular Biology. 33, 693-704.

Gillett, F.C., Forrest, W.J., and Merrill, K.M., 1973. 8 - 13-micron spectra of NGC 7027, BD +30° 3639, and NGC 6572. The Astrophysical Journal. 183, 87-93.

Glansdorff, P. and Prigogine, I., 1964. On a general evolution criterion in macroscopic physics. Physica. 30, 351-374.

Gradie, J. and Veverka, J., 1980. The composition of the Trojan asteroids. Nature. 283, 840-842.

Groves, B., Dopita, M.A., Sutherland, R.S., Kewley, L.J., Fischera, J., Leitherer, C., Brandl, B., and van Breugel, W., 2008. Modeling the pan-spectral energy distribution of starburst galaxies. IV. The controlling parameters of the starburst SED. The Astrophysical Journal Supplement Series. 176, 438-456.

Guillois, O., Nenner, I., Papoular, R., and Reynaud, C., 1996. Coal Models for the Infrared Emission Spectra of Proto-Planetary Nebulae. Astrophysical Journal. 464, 810.

Hearnshaw, J. B., 2014. The Analysis of Starlight: Two Centuries of Astronomical Spectroscopy. Cambridge University Press, New York USA, 2 edition.

Henning, T. and Salama, F., 1998. Carbon in the Universe. Science. 282, 2204–2210, 1998.

Herrmann, F. and Würfel, P., 2005. Light with nonzero chemical potential. Am. J. Phys. 73, 717-721.

Hoyle, F. and Wickramasinghe, C., 1979. On the nature of interstellar grains. Astrophysics and Space Science. 66 (1), 77-90.

Hoyle, F. and Wickramasinghe, N. C., 1999. Biofluorescence and the extended red emission in astrophysical sources. Astrophysics and Space Science. 268, 321-325.

Jewitt, D.C. and Luu, J.X., 2001. Colors and spectra of Kuiper Belt objects. The Astronomical Journal. 122, 2099-2114.

Joblin, C., Tielens, A.G.G.M., Allamandola, L.J., and Geballe, T.R., 1996. Spatial Variation of the 3.29 and 3.40 Micron Emission Bands within Reflection Nebulae and the Photochemical Evolution of Methylated Polycyclic Aromatic Hydrocarbons. Astrophysical Journal. 458, 610.

Kaifu, N., Morimoto, M., Nagane, K., Akabane, K., Iguchi, T., and Takagi, K., 1974. Detection of interstellar methylamine. The Astrophysical Journal. 191, L135-L137.

Keller, L.P., Bajt, S., Baratta, G.A., Borg, J., Bradley, J.P., Brownlee, D.E., Busemann, H., Brucato, J.R., Burchell, M., Colangeli, L., d'Hendecourt, L., Djouadi, Z., Ferrini, G., Flynn, G., Franchi, I.A., Fries, M., Grady, M.M., Graham, G.A., Grossemy, F., Kearsley, A., Matrajt, G., Nakamura-Messenger, K., Mennella, V., Nittler, L., Palumbo, M.E., Stadermann, F.J., Tsou, P., Rotundi, A., Sandford, S.A., Snead, C., Steele, A., Wooden, D., Zolensky, M., 2006. Infrared spectroscopy of comet 81P/Wild 2 samples returned by Stardust. Science. 314, 1728-1731.

Kerridge, J.F., 1999. Formation and processing of organics in the early Solar System. Space Science Reviews. 90, 275-288.

Knacke, R.F., 1977. Carbonaceous compounds in interstellar dust. Nature. 269, 132-134.

König, B.,"Organic Photochemistry", [Online] Available: http://www-oc.chemie.uni-regensburg.de/OCP/ ch/chb/oc5/Photochemie-08.pdf

Kwok, S., 2007. Molecules and solids in planetary nebulae and proto-planetary nebulae. Advances in Space Research. 40, 655–658.

Kwok, S., 2009. Organic matter in space: from star dust to the Solar System. Astrophysics and Space Science, 319, 5–21.

Kwok, S., 2012. Organic Matter in the Universe. Wiley-VCH Verlag GmbH & Co. KGaA, Weinheim Germany, p. 127-141. DOI: 10.1002/9783527637034

Kwok, S., 2016. Complex organics in space from Solar System to distant galaxies. Astronomy and Astrophysics Review. 24, 8. DOI 10.1007/s00159-016-0093-y

Kwok, S., Volk, K., and Bernath, P., 2001. On the Origin of Infrared Plateau Features in Proto-Planetary Nebulae. The Astrophysical Journal Letters. 554, L87-L90.

Kwok, S., Zhang, Y., 2011. Mixed aromatic–aliphatic organic nanoparticles as carriers of unidentified infrared emission features. Nature. 479 (7371), 80–83.

Kwok, S. and Zhang, Y., 2013. Unidentified Infrared Emission Bands: PAHs or MAONs?. The Astrophysical Journal, 771, 5. doi:10.1088/0004-637X/771/1/5

Lacy, J.H., Evans, N.J., II, Achtermann, J.M., Bruce, D.E., Arens, J.F., and Carr, J.S., 1989. Discovery of interstellar acetylene. Astrophysical Journal, Part 2 - Letters (ISSN 0004-637X). 342, L43-L46.

Leger, A. and Puget, J.L., 1984. Identification of the 'unidentified' IR emission features of interstellar dust?. Astronomy and Astrophysics. 137, L5-L8.

López-Puertas, M., Dinelli, B.M., Adriani, A., Funke, B., García-Comas, M., Moriconi, M. L., D'Aversa, E., Boersma, C., and Allamandola, L.J., 2013. Large Abundances of Polycyclic Aromatic Hydrocarbons in Titan's Upper Atmosphere. The Astrophysical Journal. 770 (2), 132.

Martins, Z., Botta, O., Fogel, M.L., Sephton, M.A., Glavin, D.P., Watson, J.S., Dworkin, J.P., Schwartz, A.W., and Ehrenfreund, P., 2008. Extraterrestrial nucleobases in the Murchison meteorite. Earth and Planetary Science Letters. 270, 130-136.

Mattila, K., Lehtinen, K., Lemke, D., 1999. Detection of widely distributed UIR band emission in the disk of NGC 891. Astronomy and Astrophysics. 342, 643-654.

McGuire, B.A., Carroll, P.B., Loomis, R.A., Finneran, I.A., Jewell, P.R., Remijan, A.J., Blake, G.A., 2016. Discovery of the interstellar chiral molecule propylene oxide ($CH_3CHCH_2O$). Science. 352 (6292), 1449-1452. DOI: 10.1126/science.aae0328

Michaelian, K. (2009) Thermodynamic origin of life. Cornell ArXiv, arXiv:0907.0042 [physics.gen-ph].

Michaelian, K., 2011. Thermodynamic dissipation theory for the origin of life. Earth System Dynamics. 2, 37–51. doi:10.5194/esd-2-37-2011

Michaelian, K., 2013. A non-linear irreversible thermodynamic perspective on organic pigment proliferation and biological evolution. J. Phys. Conference Series. 475, 012010.

Michaelian, K., 2012. HESS Opinions "Biological catalysis of the hydrological cycle: life's thermodynamic function". Hydrol. Earth Syst. Sci. 16, 2629–2645, doi:10.5194/hess-16-2629-2012.

Michaelian, K. and Simeonov, A., 2014. Fundamental molecules of life are pigments which arose and evolved to dissipate the solar spectrum. arXiv:1405.4059v2 [physics.bio-ph]

Michaelian, K. and Simeonov, A., 2015. Fundamental molecules of life are pigments which arose and co-evolved as a response to the thermodynamic imperative of dissipating the prevailing solar spectrum. Biogeosciences. 12, 4913-4937, doi:10.5194/bg-12-4913-2015.

Mumma, M.J. and Charnley, S.B., 2011. The Chemical Composition of Comets—Emerging Taxonomies and Natal Heritage. Annual Review of Astronomy and Astrophysics, 49(1), 471-524.

Nakamura-Messenger, K., Messenger, S., Keller, L.P., Clemett, S.J., and Zolensky, M.E., 2006. Organic globules in the Tagish Lake meteorite: remnants of the protosolar disk. Science. 314, 1439-1442.

NASA web site 2016, Dark Energy, Dark Matter. http://science.nasa.gov/astrophysics/focus-areas/what-is-dark-energy/ (latest access, July, 2016).

Nguyen, M.J., Raulin, F., Coll, P., Derenne, S., Szopa, C., Cernogora, G., Israël, G., and Bernard, J.M., 2007. Carbon isotopic enrichment in Titan's tholins? implications for Titan's aerosols. Planetary and Space Science. 55, 2010-2014.

Oró, J., 1960. Synthesis of adenine from ammonium cyanide. Biochemical and Biophysical Research Communications. 2, 407-412.

Oró, J. and Kimball, A.P., 1961. Synthesis of purines under possible primitive earth conditions. I. Adenine from hydrogen cyanide. Archives of Biochemistry and Biophysics. 94, 217-227.

Papoular, R., 2001. The use of kerogen data in understanding the properties and evolution of interstellar carbonaceous dust. Astronomy and Astrophysics. 378, 597-607.

Papoular, R., Conard, J., Giuliano, M., Kister, J., and Mille, G., 1989. A coal model for the carriers of the unidentified IR bands. Astronomy and Astrophysics. 217, 204-208.

Pendleton, Y.J. and Allamandola, L.J., 2002. The Organic Refractory Material in the Diffuse Interstellar Medium: Mid-Infrared Spectroscopic Constraints. The Astrophysical Journal Supplement Series. 138, 75-98.

Prigogine, I., 1967. An Introduction to the Thermodynamics of Irreversible Processes. Wiley, New York.

Puget, J.L. and Leger, A., 1989. A new component of the interstellar matter: small grains and large aromatic molecules. Annual Review of Astronomy and Astrophysics. 27, 161-198.

Roush, T.L., and Cruikshank, D.P., 2004. Observations and laboratory data of planetary organics, in Astrobiology: Future Perspectives, Ehrenfreund, P. et al. (eds.), 305, 149.

Russell, R.W., Soifer, B.T., and Willner, S.P., 1977. The 4 to 8 micron spectrum of NGC 7027. The Astrophysical Journal. 217, L149-L153.

Russell, R.W., Soifer, B.T., and Willner, S.P., 1978. The infrared spectra of CRL 618 and HD 44179 (CRL 915). The Astrophysical Journal. 220, 568-572.

Sanchez, R.A., Ferris, J.P., and Orgel, L.E., 1967. Studies in prebiotic synthesis. II. synthesis of purine precursors and amino acids from aqueous hydrogen cyanide. Journal of Molecular Biology. 30, 223-253.

Sephton, M.A., 2002. Organic compounds in carbonaceous meteorites. Natural Product Reports. 19, 292–311.

Sloan, G.C., Jura, M., Duley, W.W., Kraemer, K.E., Bernard-Salas, J., Forrest, W.J., Sargent, B., Li, A., Barry, D.J., Bohac, C.J., Watson, D.M., and Houck, J.R., 2007. The unusual hydrocarbon emission from the early carbon star HD 100764: the connection between aromatics and aliphatics. The Astrophysical Journal, 664, 1144-1153.

Smith, J.D.T., Draine, B.T., Dale, D.A., Moustakas, J., and Kennicutt Jr., R.C., 2007. The mid-infrared spectrum of star-forming galaxies: global properties of polycyclic aromatic hydrocarbon emission. The Astrophysical Journal. 656, 770-791.

Snyder, L.E. and Buhl, D., 1971. Observations of radio emission from interstellar hydrogen cyanide. The Astrophysical Journal. 163, L47-L52.

Snyder, L.E., Buhl, D., Zuckerman, B., and Palmer, P., 1969. Microwave detection of interstellar formaldehyde. Physical Review Letters. 22, 679-681.

Spencer, J.R. and Denk, T., 2010. Formation of Iapetus' Extreme Albedo Dichotomy by Exogenically Triggered Thermal Ice Migration. Science. 327 (5964), 432-435, DOI: 10.1126/science.1177132

Stocks, P.G. and Schwartz, A.W., 1981. Nitrogen-heterocyclic compounds in meteorites: significance and mechanisms of formations. Geochimica et Cosmochimica Acta. 45, 563-569.

Swings, P. and Rosenfeld, L., 1937. Considerations regarding interstellar molecules. The Astronomical Journal. 86, 483-486.

Tielens, A.G.G.M., 2005. The Physics and Chemistry of the Interstellar Medium. Cambridge University Press, New York USA, p 212.

Tielens, A.G.G.M., 2008. Interstellar polycyclic aromatic hydrocarbon molecules. Annual Review of Astronomy and Astrophysics. 46, 289-337.

Turner, B.E., Kislyakov, A.G., Liszt, H.S., Kaifu, N., 1975. Microwave detection of interstellar cyanamide. The Astrophysical Journal. 201, L149-L152.

Uchida, K.I., Sellgren, K., and Werner, M., 1998. Do the Infrared Emission Features Need Ultraviolet Excitation?. The Astrophysical Journal Letters. 493, L109.

Uchida, K.I., Sellgren, K., Werner, M.W., and Houdashelt, M.L., 2000. Infrared Space Observatory mid-infrared spectra of reflection nebulae. The Astrophysical Journal. 530, 817-833.

Vekshin, N.L., 2005. Photonics of Biopolymers, KomKniga, Moscow.

Waite Jr., J.H., 2007. The process of tholin formation in Titan's upper atmosphere. Science. 316, 870-875.

Whipple, F.L., 1978. On the Nature and Origin of Comets and their Contribution to Planets. The Moon and Planets. 19, 305-315.

Wilson, R.W., Jefferts, K.B., and Penzias, A.A., 1970. Carbon Monoxide in the Orion Nebula. Astrophysical Journal. 161, L43.

Witt, A.N. and Boroson, T.A., 1990. Spectroscopy of extended red emission in reflection nebulae. The Astrophysical Journal. 355, 182-189.